\documentclass{article}
\usepackage{spconf,amsmath,graphicx,multirow,booktabs}

\usepackage[colorlinks,
			linkcolor=black,
			anchorcolor=black,
			citecolor=black
			]{hyperref}

\usepackage{courier}

\title{CROSS-MODAL ASR POST-PROCESSING SYSTEM FOR ERROR CORRECTION AND UTTERANCE REJECTION}
\name{Jing Du, Shiliang Pu, Qinbo Dong, Chao Jin, Xin Qi, Dian Gu, Ru Wu, Hongwei Zhou}

\address{Hikvision Research Institute, Hangzhou, China}

\begin{document}
\ninept
\maketitle
\begin{abstract}
Although modern automatic speech recognition (ASR) systems can achieve high performance, they may produce errors that weaken readers’ experience and do harm to downstream tasks. To improve the accuracy and reliability of ASR hypotheses, we propose a cross-modal post-processing system for speech recognizers, which 1) fuses acoustic features and textual features from different modalities, 2) joints a confidence estimator and an error corrector in multi-task learning fashion and 3) unifies error correction and utterance rejection modules. Compared with single-modal or single-task models, our proposed system is proved to be more effective and efficient. Experiment result shows that our post-processing system leads to more than 10\% relative reduction of character error rate (CER) for both single-speaker and multi-speaker speech on our industrial ASR system, with about 1.7ms latency for each token, which ensures that extra latency introduced by post-processing is acceptable in streaming speech recognition.
\end{abstract}
\section{Introduction}
\label{sec:intro}
Modern Automatic Speech Recognition (ASR) systems can achieve high performance in terms of recognition accuracy, but they are far from perfect. The performance of ASR may degrade dramatically in mismatched domain, and errors introduced by speaker, environment and the recognizer itself will be propagated to downstream tasks. A proper post-processing system plays a complementary role for ASR system in such situation. Previous related works mainly cover five tasks, i.e. automatic error correction \cite{2020Joint}, utterance verification or rejection \cite{2000Utterance}, inverse text normalization \cite{2017A}, automatic punctuation restoration \cite{2021Replacing}, and disfluency detection \cite{2020Multi}. The former two tasks focus on improving the accuracy and reliability of ASR transcriptions, and the latter three tasks aim to achieve better readability. Since the latter three tasks and natural language understanding (NLU) tasks are all based on the result of the former two stages, error correction and utterance rejection are definitely important to avoiding error propagation in a post-processing system.

Automatic error correction is introduced into post-processing to reduce errors in ASR hypotheses, and it can be modeled in sequence tagging or sequence generation manner. Sequence tagging based methods tag each token with an editing operation, then corrected sequences are obtained by modification with these operation tags \cite{2018A}. These models may suffer from a limited set of output labels because it is difficult to cover all combinations of operations and tokens with reasonable coverage rates. Sequence generation based methods convert erroneous hypotheses to an error-free version like a machine translator \cite{2020ASR1}. In order to tackle variable length between source sentence and target sentence, some sequence generation models work in an inefficient auto-regressive way \cite{Tanaka2021CrossModalTN,2020ASR}, and some others use a length predictor to adjust the length of hypotheses \cite{2021Improved}. Neural correction models typically only use text as input \cite{2019A}. Some correction models also take acoustic features \cite{Tanaka2021CrossModalTN,2021SpellCorr} and visual features \cite{2021Read} into account, and others even infuse additional information from downstream tasks \cite{2020Joint,2020ASR}. For text-based correction task, pre-trained language model such as BERT \cite{Devlin2019BERTPO} is frequently used as backbone and shows its powerfulness and effectiveness \cite{Zhang2020SpellingEC}.

On noise corrupted speech or overlapped speech, an ASR may perform poorly, and a neural correction model also fails because of noisy context. In such cases, the cost of discarding these noisy transcriptions is much cheaper than building a high-performance recognizer, so confidence score based utterance rejection is introduced into post-processing to reject unreliable hypotheses and avoid misleading downstream tasks. Confidence score can be estimated by computing word posterior probability with lattice \cite{2000Large} or confusion network \cite{2020On} in conventional HMM-based ASR system, and it also can be approximated as softmax value of a token among all candidates \cite{2021An} for end-to-end ASR system. By such means the scores have been found over-estimated because probabilities are normalized over limited size of lattices or candidate paths. Improved estimation can be obtained by using model-based approaches, including conditional random fields (CRF) \cite{2011Combining}, multi-layer perception (MLP) \cite{2013Predicting}, recurrent neural networks (RNN) \cite{2018Speaker,2019Bi,2020Confidence}, and attention-based sequence-to-sequence models \cite{2021Confidence,2021Learning}. Various features are combined in confidence estimation. According to prior works, these features can be classified into three major categories: acoustic features (pitch, filter-bank energy, etc.), textual features (word embedding, parts of speech, etc.), and other features generated by ASR decoder (word posterior probabilities, word duration, etc.) \cite{2019System,2019Improving}.

In previous works, punctuation prediction and disfluency detection are modeled with multi-task learning \cite{Chen2020ControllableTT}. But to the best of our knowledge, none has ever tried to joint error correction and utterance rejection and explore the impact of confidence scores on correction task. In neural correction models, even if some works make progress in fusing acoustic and textual features \cite{Tanaka2021CrossModalTN,2021SpellCorr}, they use an extra speech encoder to extract acoustic representation, which is computationally expensive. 

In this paper, we train our neural correction model with a confidence estimator and a token length predictor jointly, and we find out both error corrector and confidence estimator benefit from multi-task learning. Confidence scores are used to reject unreliable hypotheses and assist neural correction model to correct erroneous tokens, which shows that confidence scores are not only critical for utterance rejection but also helpful for error correction task. Compared with single-modal model, the cross-modal model is proved to achieve higher performance even with much less parameters. The evaluation results on both single-speaker and multi-speaker speech verify the effectiveness of our proposed system.

\section{The post-processing system}
\label{sec:proposed}
\begin{figure}[h]
	\centering
	\includegraphics[scale=1.08]{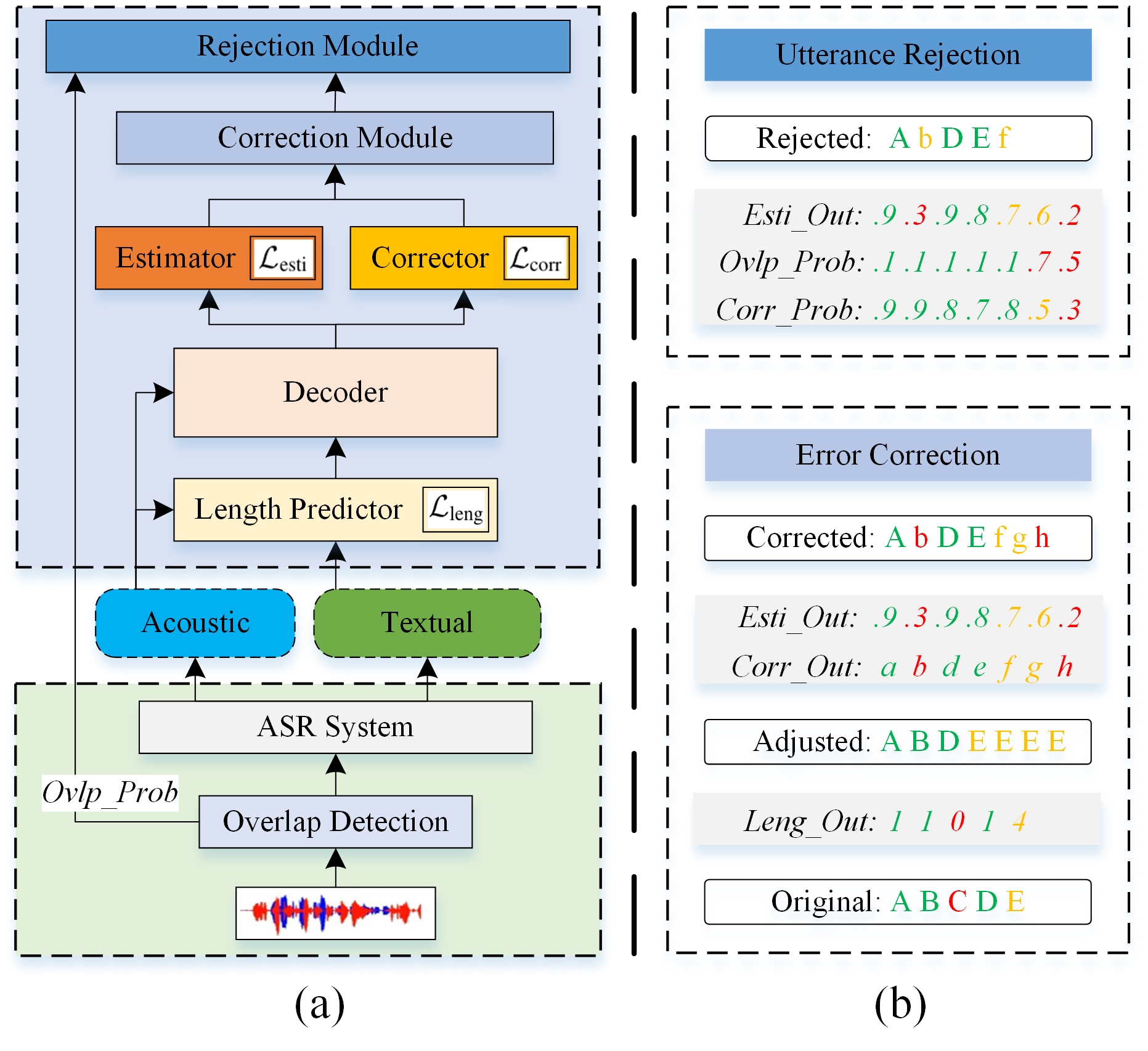}
	\caption{The post-processing system framework (a) and an example (b). Original ASR hypotheses are “A B C D E”. The length predictor adjusts them to “A B D E E E E”. The estimator outputs  confidence scores  (\textit{Esti\_Out}) and the corrector outputs a sequence “a b d e f g h” (\textit{Corr\_Out}). Tokens with high confidence scores are recovered, so we get corrected sequence “A b D E f g h”. According to confidence scores, overlap probabilities (\textit{Ovlp\_Prob}), and sofmax values of the corrector output (\textit{Corr\_Prob}), we reserve the token “b” (although its confidence score is only 0.3, it may be corrected because the corrector gives a high softmax value 0.9), and we reject “g” and “h”.}
	\label{fig:framework}
\end{figure}

Figure~\ref{fig:framework} illustrates the framework of our proposed system, and it also gives an example of error correction and utterance rejection.  
\subsection{Cross-modal features}
\label{ssec:features} 
As prior post-processing models, we use ASR one-best hypotheses as pre-requisite input. We also introduce some features from WFST-based decoder, such as word probabilities and duration in lattice. Characters in ASR output text are mapped into continuous space representation by an embedding layer. Word probabilities and duration from ASR decoder are converted into character-level then concatenated to text embedding, by which means we get our textual-modal features. 

Furthermore, we fetch hidden states of acoustic model (AM) in ASR system as our acoustic-modal features, because the hidden states of AM can be seen as a bottleneck feature representing high-level speech embedding. Compared with using raw speech or filter-bank energies, using hidden states of AM eliminates an extra speech encoder for extracting acoustic representation.
\subsection{Multi-task learning}
\label{ssec:multi-task} 
We use a multi-task loss function to train our model. Since we joint three tasks, our loss tuple contains three task-specific cross entropy (CE) loss elements, i.e. length predictor loss \(\mathcal L_{\text {leng}}\), confidence estimator loss \(\mathcal L_{\text {esti}}\), and error corrector loss \(\mathcal L_{\text {corr}}\). In our experiments, we verify that language model distillation loss \(\mathcal L_{\text {dist}}\) and confidence scores based regularization loss \(\mathcal L_{\text {conf}}\) are helpful for our decoder. The total loss is given by
\begin{equation}
	\label{eq_loss}
\mathcal L_{\text{total}}=\mathcal L_{\text {leng}} +\mathcal L_{\text {esti}} + \mathcal L_{\text {corr}} + \mathcal L_{\text {dist}} + \mathcal L_{\text {conf}}.
\end{equation}

The \(\mathcal L_{\text {dist}}\) can be obtained by calculating mean square error (MSE) between the output of our decoder and a pre-trained language model, and the \(\mathcal L_{\text {conf}}\) is defined as
\begin{equation}
	\label{eq0}
\mathcal L_{\text {conf}}=(1-c) /(1-\bar{c}) \operatorname{KLDiv}\left(o_{\text {corr}}, y\right),
\end{equation}
where c represents the confidence scores output by the estimator in a mini-batch, \(\bar{c}\) is the mean confidence score of the mini-batch, and \(\operatorname{KLDiv}\left(o_{\text {corr}}, y\right)\) is Kullback-Leibler divergence between the output of our corrector and the ground truth. The item \((1-c) /(1-\bar{c})\) can be seen as a weight for per token, the larger the confidence score, the smaller the weight. By such regularization, unreliable tokens and sequences in a mini-batch are given larger penalties, so that the corrector can learn the difficult samples sufficiently.

\subsection{Error correction}
\label{ssec:joint} 
In correction module, original hypotheses are first adjusted with the prediction of token length (\textit{Leng\_Out}). We adopt a similar method as FastCorrect \cite{Leng2021FastCorrectFE}: a token should be removed if its length prediction value is 0, and if its length \(x\) is more than 1, it should be duplicated by \((x-1)\) times, otherwise the token should remain unchanged. 

The confidence estimator is a binary classifier and it outputs the confidence score of a token (\textit{Esti\_Out}). The corrector produces many candidates for each token, and we simply choose the token with maximum softmax value as its output (\textit{Corr\_Out}). If the confidence score of a token is lower than a specific threshold, this token is substituted by the corrector's output, on the contrary, if its confidence score is high enough, the corrector's output of this token will be neglected. We call this approach as “confidence scores filtering”, with which we can avoid modifying original correct tokens by mistake.
\subsection{Utterance rejection}
\label{ssec:rejection} 
To reject hypotheses transcribed from overlapped speech, we train a three-class voice activity detection (VAD) model called “overlap detection” to detect whether a frame of speech lies in overlapped part or not. The overlap detection model may output 0, 1, 2, representing silence/noise, single-speaker speech, and overlapped speech, respectively. We convert probabilities of the third class (labelled with 2) from frame-level to character-level to represent overlap probabilities (\textit{Ovlp\_Prob}) of each character in hypotheses.

In rejection module, each token is judged by a three-element tuple containing confidence score, overlap probability, and maximum softmax value of the corrector's output (called corrector probability, \textit{Corr\_Prob}). Tokens with low confidence scores, high overlap probabilities, and low corrector probabilities are removed, because they are either unreliable, or in overlapped speech, and the corrector are not confident to correct these tokens. The rejection can be performed in different levels, such as token level, segment level, and sequence level. To achieve better performance, confidence scores may be preprocessed by some smoothing algorithms before rejection. 
\section{Experiments}
\label{sec:Experiments}
\subsection{Baseline model}
\label{ssec:baseline}
Our ASR system is based on feedforward sequential memory networks (FSMN) \cite{2018Deep}. We train an acoustic model with 24k-hour Mandarin speech data in CE criteria followed with MMI training \cite{K2013Sequence}. The acoustic features are 80-dimensional log-mel filter-bank energies computed on 25ms window with 10ms shift, which are down-sampled by 3 times for low-frame-rate training and decoding. And our 4-gram language model is trained with about 800G corpus. This ASR system yields 3.3\% CER in Aishell-1 test set \cite{2017AISHELL}.

We fine-tune a pre-trained BERT model known as bert-base-chinese$\footnote{https://huggingface.co/bert-base-chinese}$ to perform confidence estimation and error correction. The BERT output layer is followed with two parallel linear layers, which are used as confidence estimator (output size is 2) and error corrector (output size is 5200). This baseline model only uses ASR transcription as input.
\subsection{Post-processing model}
\label{ssec:post-process}
We train our post-processing model with ESPnet toolkit \cite{2018ESPnet}. Our model is based on standard transformer blocks. The dimension of token embedding, hidden state and the feedforward layers are 256, 256 and 2048, respectively, and the attention head size is 4. Because we use a deep layer of AM to export acoustic embedding, we can minimize the post-processing model's acoustic encoder with just one linear layer to convert the dimension of acoustic embedding from 512 to the decoder’s hidden state size. 

The length predictor consists of 2 transformer blocks followed with a linear layer. In consideration of that too many consecutive deletion tokens are difficult to recover, we set the linear output size of the length predictor as 5, which means maximum consecutive number of deletion errors we can model is 3. The decoder consists of 6 transformer blocks, followed with two parallel linear layers: one act as confidence estimator and the other as error corrector.

In order to avoid the confidence estimator becoming over confident on ASR training data, we choose about 1/4 of total ASR training data, whose recognition accuracy is comparatively lower than average, to train our post-processing models.

\subsection{Evaluation settings}
\label{ssec:setting}
\subsubsection{Evaluation data}
\label{sssec:data}	
The single-speaker test sets include a 180-hour in-domain close-talk (\texttt{IDCT}) dataset that consists of 18 subsets, a 10-hour out-of-domain close-talk (\texttt{ODCT}) dataset whose context involves lyric, poetry, chitchat and so on, and a 20-hour out-of-domain far-field (\texttt{ODFF}) dataset. The multi-speaker test sets contain two 5-hour subsets. One is simulated by mixing 2 single-speaker audio with specific proportions of duration and volume (\texttt{SMS}), and the other is collected from real world (\texttt{CMS}). It is worth noting that the overlapped part of multi-speaker test audio is not transcribed, so if the ASR hypotheses lie in overlapped speech, they will be treated as insertion errors. We want to evaluate how many insertion errors our utterance rejection module can remove correctly for overlapped speech.
\subsubsection{Evaluation metric}
\label{sssec:metric} 
We use areas under the curve (AUC) and normalized cross entropy (NCE) \cite{1997Improved} to measure confidence scores. Receiver operating curve (ROC) is used to illustrate the operating characteristic in our evaluation, which is a curve of false positive rate (FPR) against true positive rate (TPR) in a range of threshold values. Since different confidence score distributions may have the same AUC value, we need NCE to evaluate the confidence estimator. For a given test set, let  \({\mathbf{s}=\left[s_{1}, s_{2}, \ldots, s_{N}\right]}\)  and \(\mathbf{\ell}=\left[l_{1}, l_{2}, \ldots, l_{N}\right]\)  represent confidence scores and labels for total N tokens, respectively, where \(\mathbf{s} \in[0,1]\)  and \(\mathbf{\ell} \in\{0,1\}\) . We use \(H(\mathbf{\ell})\)  as entropy of target label sequence and \(H(\mathbf{s} \mid \mathbf{\ell})\)  as the cross-entropy between distributions of confidence scores and labels. The NCE is given by 
\begin{equation}
	\label{eq3}
	\mathrm{NCE}=\frac{H(\mathbf{\ell})-H(\mathbf{s} \mid \mathbf{\ell})}{H(\mathbf{\ell})}.
\end{equation}

The performance of estimator is better when NCE value is larger. If the confidence scores perfectly match the token labels, the NCE value would equal to 1.

We use CER to measure the performance of our error correction and utterance rejection modules. For rejection task, CER reduction represents positive net rejection, which means total numbers of true rejections is more than false rejections. This is the reason why we use CER instead of false rejection rate to evaluate our utterance rejection module.

\subsection{Evaluation results}
\label{ssec:results}
We first evaluate the performance of our confidence estimator. The result is showed in Table~\ref{tab:confidence and correction}. As we can see, NCE values of lattice-based confidence scores given by ASR system are all negative, which reflects that using word probability in lattice as confidence score is unreliable. Pre-trained BERT offers a competitive baseline after fine-tuning, and our estimator gives obviously better scores than BERT on all test sets. For multi-speaker speech, tokens may be transcribed correctly and have high confidence scores even if they lie in overlapped speech, but they are labelled with “0” actually. This is the reason why both BERT and our estimator perform poorly on multi-speaker test sets.

\begin{figure}[h]
	\centering
	\includegraphics[scale=0.48]{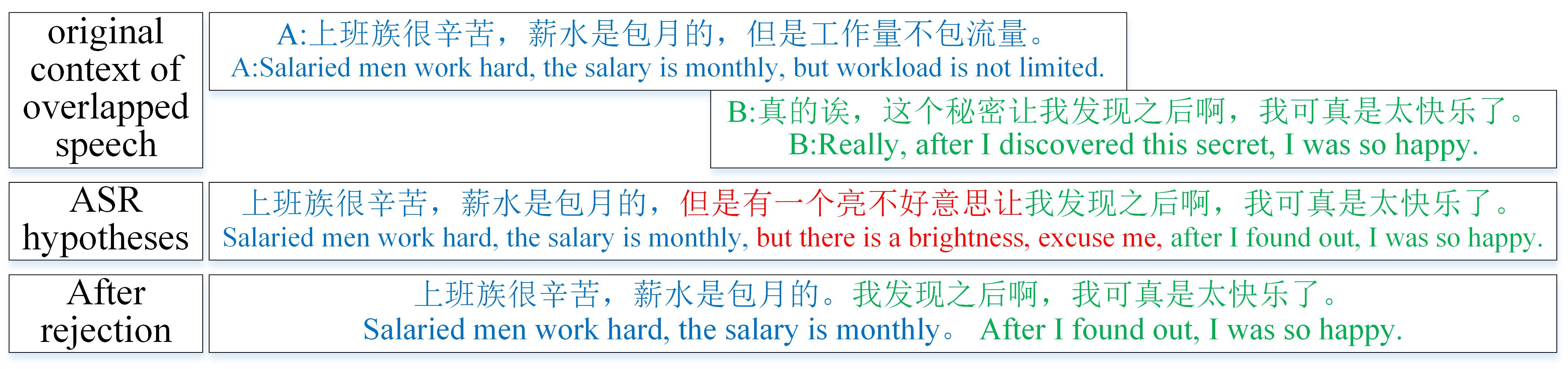}
	\caption{An example of the utterance rejection result. Text in blue and green are from different speakers. ASR outputs some incorrect and confusing words in overlapped speech with red color. After rejection, more reliable and readable context is left.}
	\label{fig:rejection}
\end{figure}

Table~\ref{tab:confidence and correction} also shows the effect of our error correction and utterance rejection modules. For single-speaker speech, BERT baseline results in about 6\% relative character error rate reduction (CERR). Our error correction module outperforms BERT with 13\% relative CERR on \texttt{IDCT}, and more than 8\% relative CERR on \texttt{ODCT} and \texttt{ODFF}, which proves the effectiveness and robustness of our corrector. After we reject some extremely unreliable tokens, the CER decreases further. On multi-speaker sets, both BERT and our corrector fails because most of errors are insertion errors in overlapped speech. With our rejection module, we can delete these inserted tokens. Result shows that our rejection module leads to more than 12\% relative CERR on \texttt{SMS} and \texttt{CMS}. It’s worth noting that because nearly half of the audio is overlapped in multi-speaker test sets, even if we reject some inserted tokens, the CER is still high.  An example of overlapped speech transcription rejection is given by Figure~\ref{fig:rejection}, which proves that even though we don’t recognize overlapped speech correctly, we can improve the reliability of the transcription by our rejection module. 
\begin{table*}[h]
	\caption{Comparison of confidence measurement and CER among ASR , BERT baseline and our system on different test sets}
    \vspace{10pt}
	\label{tab:confidence and correction}	
	\centering
	\setlength{\tabcolsep}{1.mm}
	{
		\begin{tabular}{cccccccccccccccc}
			\toprule[1pt]
			\multirow{2}{*}{} & \multicolumn{3}{c}{\texttt{IDCT}} & \multicolumn{3}{c}{\texttt{ODCT}} & \multicolumn{3}{c}{\texttt{ODFF}} & \multicolumn{3}{c}{\texttt{SMS}} & \multicolumn{3}{c}{\texttt{CMS}} \\
			& AUC    & NCE   & CER(\%) & AUC    & NCE   & CER(\%) & AUC    & NCE   & CER(\%) & AUC   & NCE   & CER(\%) & AUC   & NCE   & CER(\%) \\
			\midrule
			ASR      & 0.856  & -0.28 & 11.53   & 0.834  & -0.34 & 12.41   & 0.827  & -0.43 & 17.21   & 0.775 & -1.93 & 41.4    & 0.76  & -2.19 & 45.22   \\
			BERT baseline	&0.887	&0.248	&10.75	&0.859	&0.235	&11.73	&0.85	&0.223	&16.35	&0.783	&0.102	&41.3	&0.779	&0.8	&45.06 \\			
			our estimator     & 0.912  & 0.316 & -       & 0.882  & 0.307 & -       & 0.879  & 0.302 & -       & 0.791 & 0.116 & -       & 0.783 & 0.103 & -     \\
			\quad+correction     & -      & -     & 10.00   & -      & -     & 11.39   & -      & -     & 15.55   & -     & -     & 41.26   & -     & -     & 44.98  \\
			\quad\quad+rejection   & -      & -     & 9.89    & -      & -     & 11.34   & -      & -     & 15.46   & -     & -     & 35.13   & -     & -     & 39.37   \\
			\bottomrule[1pt]    
		\end{tabular}
	}
\end{table*}
\section{Analysis}
\label{sec:Analysis}
\subsection{Cross-modal model works}
\label{ssec:backbones}
To explore the impact of different features, we fuse 4 types of features in stages. The result is showed in Table~\ref{tab:features}. As we can see, text-based BERT is powerful and performs much better than our text-only model. To our surprise, after fusing acoustic embedding with textual feature, our model beats BERT on confidence estimation task, and achieves nearly the same CER on error correction task with 8.2 times inference acceleration and just 1.7ms latency for each token, which proves the importance of cross-modal features in these tasks. By adding more features from our WFST-based decoder, the performance improves further.
\begin{table}[h]
	\caption{Performance of different backbones with different features on \texttt{IDCT}: \textit{text}, \textit{acou}, \textit{prob}, and \textit{dur} represent ASR hypotheses, acoustic embedding from AM's hidden states, lattice-based word probabilities and duration for each token, respectively.}
	\vspace{10pt}
	\label{tab:features}
	\centering
	\setlength{\tabcolsep}{1.2mm}
	{
		\begin{tabular}{cccccl}
			\toprule[1pt]
			{Backbones} & Features & {AUC} & {NCE} & {CER}(\%) & \begin{tabular}[c]{@{}c@{}}Inference\\Speed\end{tabular} \\
			\midrule
			BERT                 & \textit{  text}     & 0.887 & 0.248 & 10.75   & \quad \(\times\)1            \\
			Joint model     & \textit{  text}     & 0.859 & 0.106 & 11.34   & \quad \(\times\)13.1           \\
			& \quad\textit{+acou}    & 0.892 & 0.276 & 10.77   & \quad \(\times\)8.2             \\
			& \quad\quad\textit{+prob}    & 0.903 & 0.288 & 10.67   & \quad \(\times\)8.2             \\
			& \quad\quad\quad\textit{+dur}     & 0.906 & 0.292 & 10.6   & \quad \(\times\)8.2   \\
			\bottomrule[1pt]
	\end{tabular}}
\end{table}
\subsection{Multi-task learning matters}
\label{ssec:matters}
In Table~\ref{tab:joint}, we explicitly compare our joint model with separate confidence estimator and error corrector. Two single models and our joint model use the same structure and model size. By multi-task training, the performance improves in both tasks, and total model parameters is halved, which improves the efficiency too. The result proves that the confidence estimation task and error correction task illuminate each other. The estimator acts like an error detector to guide our corrector in multi-task training, and the corrector also help the estimator to discovery erroneous tokens. 
\begin{table}[h]
	\caption{Comparison between single estimator/corrector and our joint model, and results of other ablation studies on \texttt{IDCT}.}
	\vspace{10pt}
	\label{tab:joint}
	\setlength{\tabcolsep}{1.5mm}
	\centering
	\begin{tabular}{l c c c c}
		\toprule[1pt]
		& {AUC} & {NCE} & {CER(\%)} & Size(MB)  \\
		\midrule
		Estimator   only      & 0.894        & 0.26         & -    & \quad 46.8             \\
		Corrector   only      & -            & -            & 10.91   &\quad 46.8          \\
		\text{\quad+confidence filtering} & -            & -            & 10.7        &\quad 46.8\(\times\)2      \\
		Joint   model     & 0.906        & 0.292        & 10.6         &\quad 46.8     \\
		\quad+confidence filtering & 0.906        & 0.292        & 10.55        &\quad 46.8     \\
		\quad\quad+length adjustment    & 0.907        & 0.293        & 10.35       &\quad 62.4      \\
		\quad\quad\quad+regularization       & 0.909        & 0.302        & 10.2         &\quad 62.4     \\
		\quad\quad\quad\quad+distillation         & 0.912        & 0.316        & 10.0     &\quad 62.4         \\
		\bottomrule[1pt]
	\end{tabular}
\end{table}

There is another comparison between with and without “confidence filtering”. The term “with confidence filtering” means that our corrector only modifies tokens whose confidence scores are under a specific threshold. The result shows that if we use corrector with confidence scores filtering, the single corrector gains 0.21 CERR, but our joint model only gains 0.05 CERR. This result reflects the impact of confidence score on correction task. Since the corrector of joint model has already been influenced by the estimator implicitly in multi-task training, the explicit confidence filtering is nearly useless.
\subsection{Other ablation studies}
\label{ssec:ablation}

In this section, we introduce some of our optimizing techniques in model training. The result is showed in Table~\ref{tab:joint}. As the result showing, CER decreases by 0.2 when we use our length predictor to adjust the length, which proves the effectiveness of the length predictor in error correction task. According to the statistics of different error types, we find out that the length predictor mainly contributes to reduction of insertion errors. After we use an additional regularization loss based on confidence scores explained in section~\ref{ssec:multi-task}, both the estimator and the corrector are enhanced, which implies the difficult training examples may be learnt more sufficiently with our regularization loss. We also use pre-trained bert-base-chinese as a language model to distill our decoder, by which means we believe external knowledge is introduced into our decoder. The result also verifies our guess. The performance of both estimator and corrector improve after distilled by a language model.
\section{Conclusions}
\label{sec:conclusions}
In this paper, a cross-modal ASR post-processing system unifying error correction and utterance rejection is built with multi-task learning, and experiments prove its high performance and high computational efficiency. With such system, we can correct erroneous tokens and filter out unreliable tokens, so that more accurate and reliable results are conveyed to human readers and downstream applications. In the future, we will focus on modeling utterance rejection with neural networks instead of using logical judgment, to make the post-processing system more convenient and robust.


\vfill\pagebreak


\bibliographystyle{IEEEbib}
\bibliography{my_paper}

\end{document}